\documentclass[twocolumn,showpacs,preprintnumbers,pra,amsmath,amssymb]{revtex4}

\usepackage{graphicx}
\usepackage{dcolumn}
\usepackage{bm}
\usepackage{amsmath}
\usepackage{amssymb,amsthm}
\usepackage{epsfig,color}
\usepackage[sans]{dsfont}

\definecolor{nblue}{rgb}{0.2,0.2,0.7}
\definecolor{ngreen}{rgb}{0.2,0.6,0.2}
\definecolor{nred}{rgb}{0.7,0.2,0.2}
\definecolor{nblack}{rgb}{0,0,0}

\newcommand{\tr}{\text{tr}}

\newcommand{\Id}{\mathds{1}}

\def\E{\mathcal{E}}

\newcommand{\C}{\mathbb{C}}

\def\tr{\mbox{tr}}

\def\bea{\begin{eqnarray}}
\def\eea{\end{eqnarray}}
\begin{document}

\title{Superactivation, unlockability, and secrecy distribution of bound information}

\author{Giuseppe Prettico$^{1}$ and Joonwoo Bae$^{2}$}

\affiliation{$^{1}$ICFO-Institut de Ciencies Fotoniques, E-08860 Castelldefels, Barcelona, Spain \\
$^{2}$School of Computational Sciences, Korea Institute for
Advanced Study, Seoul 130-722, Korea}

\date{\today}

\begin{abstract}
Bound information, a cryptographic classical analogue of bound entanglement, is defined as classical secret correlations from which no secret key can be extracted. Its existence was conjectured and shown in a multipartite case. In this work, we provide a new example of bound information in a four-partite scenario. Later, using this example, we prove that bound information can be superactivated in a finite-copy scenario and unlockable. We also show that bound entangled states (bound information) can be used to distribute multipartite pure-state entanglement (secret key).
\end{abstract}

\maketitle

\section{Introduction}

Entanglement is one of the key resources that distinguish Quantum
Information Theory from its classical counterpart. The
impossibility for spatially separated parties to create entangled
states by local operation and classical communication (LOCC)
underlies the crucial role that entanglement plays for
communication purposes. The property that entanglement is
non-increasing under LOCC naturally leads to the definition of the
maximally entangled state \cite{ref:unit}, \bea|\psi_{AB}\rangle =
( |00 \rangle + |11\rangle) / \sqrt{2}. \nonumber \eea This state
is often called an entangled bit (ebit) and used as the unit of
entanglement. Distilling ebits is a basic task in quantum
communication scenarios.

The quantum distillation scenario shares many similarities with the information-theoretic secret key distillation scenario introduced in Ref. \cite{ref:keyagreement}. In this classical scenario, a secret bit (sbit) is aimed to be distilled from initially shared secret correlations using local operations and public communication (LOPC). An sbit is described by a probability distribution  among honest parties Alice (A) and Bob (B) and an eavesdropper Eve ($\E$), $P_{AB\E}$, satisfying: \bea P_{AB\E} (a,b,e) & = & P_{AB}(a,b) ~P_{\E}(e), \nonumber \\ P_{AB}(a,b) & = & \delta_{a,b}/2\label{eq:sbit} \eea for $a,b\in \{0,1\}$. The former condition means that the honest parties are uncorrelated with Eve, and the latter shows perfect correlations between them. Analogous to properties of entanglement, secret correlations do not increase under LOPC \cite{ref:Analogy}. Therefore, an sbit represents the maximal secret correlations and is used as the unit of classical correlations.

One can then relate an ebit with an sbit and vice versa as the basic analogues of units of correlations. Based on this correspondence, further analogies between quantum and classical scenarios follow, e.g. LOPC as a
classical analogue of LOCC \cite{ref:Analogy}. Moreover, interrelations of information-theoretic quantities and their operational meanings have also been
investigated in \cite{ref:linking} \cite{ref:AnalogyCerf} \cite{ref:squash}.\\

There are also well-established quantitative relations between quantum and secret correlations. To introduce them, let us briefly remind two entanglement measures presented in Refs \cite{ref:entcost, ref:distillent}. Each measure has operational meaning, one for the formation and the other for the distillation of enangled quantum states. The entanglement cost denoted by $E_{c}$ quantifies the number of ebits for the formation of a given quantum state $\rho_{AB}$ in the asymptotic limit, i.e. when the number of copies of $\rho_{AB}$ tends to be very large \cite{ref:entcost}. Entanglement cost is positive if and only if a given state is entangled \cite{ref:H3}. In a similar way, the number of ebits that can be distilled out of given quantum states in the asymptotic limit is quantified and called entanglement of distillation denoted by $E_{D}$ \cite{ref:distillent}. In general, it holds that $E_{c} \geq E_{D}$ as one cannot get more entanglement from given states than that used for the preparation. The existence of undistillable, i.e. bound, entangled states \cite{ref:boundent}, shows the irreversibility in the entanglement manipulation and appears when $E_{c} > E_{D} =0$. It has been shown that, despite their undistillability \emph{per se}, bound entangled states can be activated \cite{ref:boundent} \cite{ref:EntAct} \cite{ref:durciracjpa}.

Similar questions can be addressed in the context of the classical cryptographic scenario. That is, when a probability distribution $P_{AB\E}$ is given, the information of formation denoted by $I_{c}$ has been derived as the classical analog of $E_{c}$ \cite{ref:inf-form}, and determines the number of sbits for the formation of a given distribution via LOPC. Positive information of formation, $I_{c} (A : B | C) >0$, means that a given probability distribution contains secret correlations. For the distillation, the natural classical analog is the secret key rate, denoted by $S(A:B \| \E)$ \cite{ref:keyrate}, the number of sbits that can be distilled from given classical correlations in the asymptotic limit.

Quantum and classical correlations can be related by measurement, i.e. $P_{AB\E}= \tr[\rho_{AB\E} M_{A}\otimes M_{B}\otimes M_{\E}]$ where $M_{j}$ for $j=A,B,\E$ are positive-operator-valued-measure on quantum state $\rho_{AB\E}$. It immediately follows that an sbit is obtained by measuring an ebit in the computational basis. Entanglement and secret correlations are also generally interrelated. That is, given quantum states, if they are entangled, there always exist local measurements such that the measured outcomes contain secret correlations. As well, given probability distributions obtained by measuring quantum states, if they consist of secret correlations, then the corresponding quantum states must be entangled, i.e. $E_{c}>0 \Leftrightarrow I_{c}>0$ \cite{ref:SecCor}.

As a classical analog of bound entanglement, bound information was defined as secret correlations from which no sbit can be distilled \cite{ref:linking}. The existence of bound information was then explored, and was indeed shown in Ref. \cite{ref:multibouninf} . An instance of bound information was explicitly provided in a multipartite scenario, and remarkably, was obtained by measuring bound entangled states presented in Ref. \cite{ref:durciracjpa}. Moreover, it was shown that bound information can be activated in an asymptotic-copy scenario \cite{ref:multibouninf}. Note that this is also called superactivation in the sense that individual probability distributions consisting of bound information are undistillable while the one combined together can be converted to distillable correlations.

In this work, along the interrelation between quantum and secret correlations, we study further properties of bound information. We introduce the classical analogues of finite-copy superactivation and the unlockability of bound entanglement. All these findings are based on the intriguing properties of the Smolin state introduced in Ref. \cite{ref:SmolinStates}. We also discover a useful feature of the undistillable correlations in distributing pure-state entanglement and multipartite sbits. In the quantum scenario, it is shown that the tripartite GHZ state can be extended to the four-partite GHZ state using LOCC, given that a four-partite bound entangled state
is shared among parties. A classical analogue also follows. When bound information is shared by four parties, an sbit of three parties can be distributed over the four parties using LOPC. Note also that the Smolin state is immediately a feasible resource that has been implemented with present-day technology \cite{ref:exp}.

This paper is organized as follows. In Sec. \ref{sec:smolin state}, the properties of Smolin states are briefly reviewed. In Sec. \ref{sec:boundinf}, the bound information is then derived by measurement on the Smolin state, and the properties such as unlockability and superactivation are translated. In Sec. \ref{sec:dis}, it is shown that bound entangled state (bound information) together with LOCC (LOPC) can be used to extend GHZ states (sbits) from three to four parties.

\section{The Smolin State}
\label{sec:smolin state}

Let us first briefly review the properties of the Smolin state presented in Ref. \cite{ref:SmolinStates}. The Smolin state is a four-partite bound entangled state, shared by, say Alice, Bob, Clare and David: \bea \rho_{ABCD} = \frac{1}{4} \sum_{i} |\psi_{i}\rangle_{AB}\langle \psi_{i}|\otimes |\psi_{i}\rangle_{CD}\langle \psi_{i}|, \label{eq:smolin}\eea where $|\psi_{1}\rangle = (|00\rangle+|11\rangle)/\sqrt{2}$, $ |\psi_{2}\rangle = (|00\rangle - |11\rangle)/\sqrt{2}$, $|\psi_{3}\rangle = (|01\rangle+|10\rangle)/\sqrt{2}$, and $|\psi_{4}\rangle = (|01\rangle - |10\rangle)/\sqrt{2}$. This state has been exploited to derive intriguing effects of bound entanglement such as the unlockability and the superactivation in a finite-copy scenario \cite{ref:SmolinStates}. Let us summarize the properties in the following.

\begin{itemize}
  \item (i) \emph{Invariance under permutations.} The state is symmetric under any exchange of parties, i.e. $\rho_{ABCD} = \rho_{ABDC} = \rho_{ADBC}$.

  \item (ii) \emph{Undistillability.} Looking at the bipartite splitting $AB:CD$ in the state in Eq. \eqref{eq:smolin}, it is clear that the state is separable across the cut. Then, from the property (i), it follows that the state is separable in all bipartitions across two parties versus the others, such as $AC:BD$ and $AD:BC$. This already shows that no pair of parties can distill entanglement, and therefore the state is undistillable.

  \item (iii) \emph{Unlockability.} An important property of the state is the unlockability of entanglement. This can be seen when two parties among the four join together and apply collective operations to discriminate among the four Bell states. Announcing the measurement outcome, the two joined parties can allow the other two parties to know which Bell state is shared between them. Then, applying local unitaries that depend on the announced outcome, they can finally distill the Bell state $|\psi_{1}\rangle$.  This shows that the Smolin state is entangled, and also bound entangled together with the property (ii).
\end{itemize}

\begin{figure}
\begin{center}  
  \includegraphics[width= 9.2 cm]{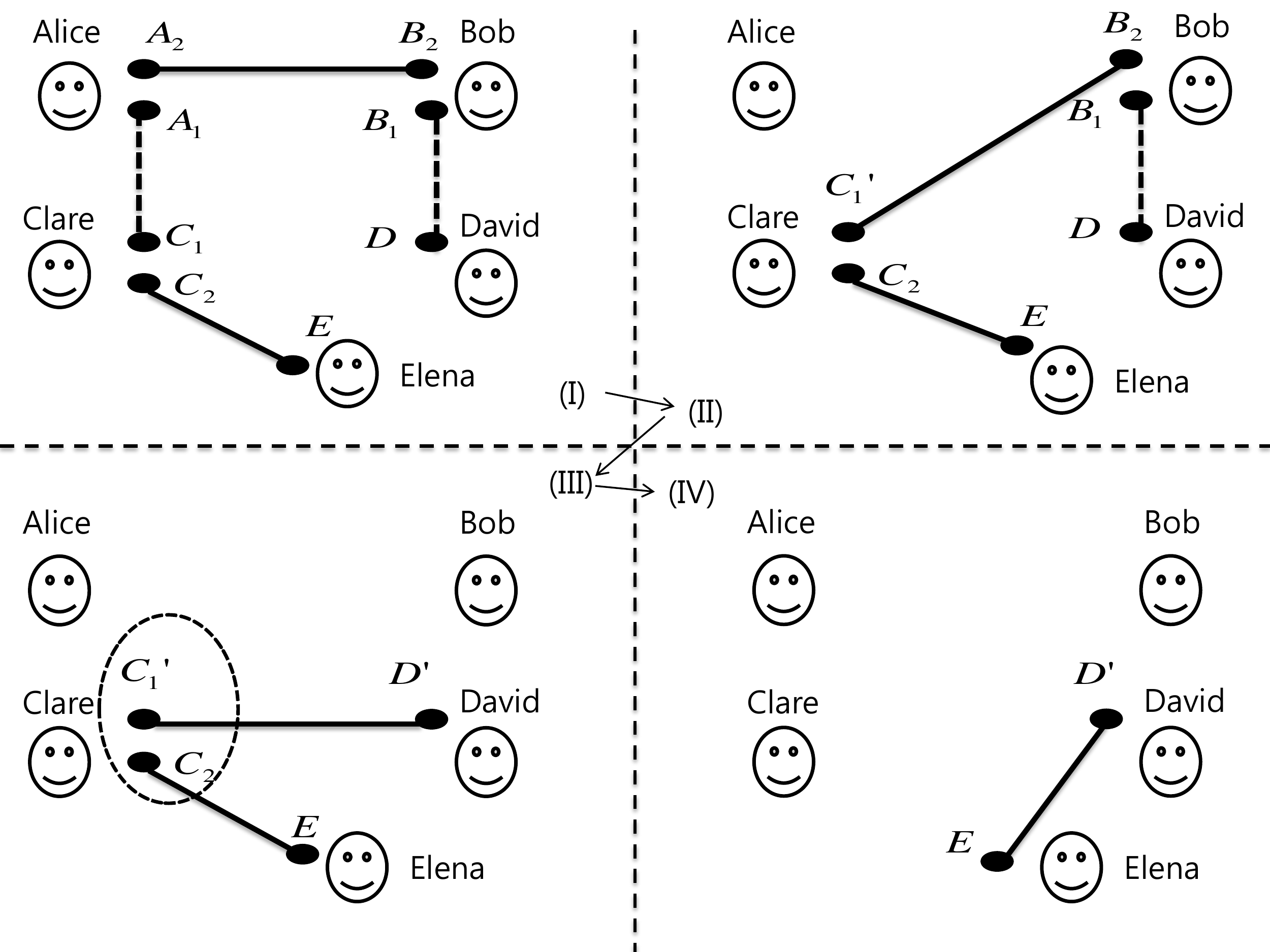}
  \caption{The activation protocol for both the quantum and classical scenarios is shown. For the quantum scenario, two Smolin states $\rho_{A_{1}C_{1}B_{1}D}$ and $\rho_{A_{2}B_{2}C_{2}E}$ are drawn with the dashed and solid lines, respectively. For the classical scenario, the Smolin states are simply replaced with the bound information in (\ref{smolin}). In both cases, the first step in the protocol (shown in I$\rightarrow$II) is that Alice teleports her state in the system $A_{2}$ to $C_{1}$ using the correlation between $A_{1}C_{1}$. In the second step (shown in II$\rightarrow$III), Bob teleports the state of his system $B_{2}$ to $D$ sacrificing the correlation existing in $B_{1}D$. Then, the resulting distribution over the remaining three parties, Clare holding two systems, David, and Elena is in fact the Smolin state if the scenario is with quantum systems, or the bound information in (\ref{smolin}) if it is with classical systems. Finally (shown in III$\rightarrow$IV
 ), Clare measures her systems and announces the outcomes, so that David and Elena distill an ebit or an sbit.}\label{fig1}
\end{center}
\end{figure}

\emph{Superactivation with finite copies.} One of the intriguing effects in the entanglement theory is that bound entangled states can be activated. The Smolin state was in particular exploited to show a strong version of the effect, superactivation in a finite-copy scenario. It is called superactivation in the sense
that $E_{D}(\rho_{1}\otimes\rho_{2}) > 0$ while $E_{D}(\rho_{1}) = E_{D}(\rho_{2} ) = 0$. This was shown in Ref. \cite{ref:SuperActivation}, and the activation works as follows. Suppose that, now including the fifth one Elena, two copies of the Smolin state are shared by the five parties in the following way,  \bea \rho_{A_{1}C_{1}B_{1}D} \otimes \rho_{A_{2}B_{2}C_{2}E}, \label{eq:DEactivation} \eea where the first and the second copies are labeled. Then, David and Elena
distill an ebit, applying the following protocol, see also Fig. \ref{fig1}. First, Alice teleports her qubit state of $A_{2}$ to Clare sacrificing the unknown Bell state shared between $A_{1}$ and $C_{1}$. Clare is then with two qubits $C_{1}^{'}$ and $C_{2}$ where $C_{1}^{'}$ is in the teleported state from $A_{2}$. Next, Bob teleports his qubit state of $B_{2}$ to $D$ using the unknown Bell state shared between $B_{1}$ and $D$. Then, David is now with $D^{'}$ in the teleported state from $B_{2}$. Finally, due to the structure of the Smolin state, the state $C_{1}^{'}C_{2}D^{'}E$ shared by Clare, David, and Elena results in the Smolin state. Since Clare holds two qubits $C_{1}^{'}$ and $C_{2}$, she can discriminate among Bell states and announces the result, by which David and Elena can distill an ebit. Now, symmetrizing the state in Eq. (\ref{eq:DEactivation}) with more copies as follows, \bea \rho_{ABCD} \otimes \rho_{ABCE} \otimes \rho_{ABDE} \otimes \rho_{ACDE} \otimes \rho_{BCDE}, \label{eq:tensoring} \eea any two parties among the five can distill ebits. In this way, from ebits shared by every two parties, it follows that multipartite pure entangled states can be distilled. This finally shows that multipartite bound entangled states can be superactivated. It is also noteworthy in the activation scenario that, the distillable entanglement defined in the asymptotic limit becomes immediately positive in a finite number of copies.

\section{Bound information}
\label{sec:boundinf}

Bound information is characterized by two constraints on probability distributions, i) existence of secret correlations $I_{c}(A:B|\E) > 0$ and ii) the undistillability $S(A:B\|\E) =0$. There is an information measure located in the between, the intrinsic information $I(A:B \downarrow \E)$ \cite{ref:keyagreement}, such that \bea I_{c}(A:B|\E)\geq I(A:B \downarrow  \E) \geq S(A:B\|\E). \label{eq:intinq} \eea Its usefulness lies on the fact that a general undistillability criterion is lacked and zero-valued intrinsic information is immediately a sufficient condition for the undistillability. This condition is therefore going to be used later on, when proving that a given probability distribution contains bound information. The intrinsic
information is defined as the minimum of the conditional mutual information of honest parties given an eavesdropper, \bea I(A:B \downarrow  \E) = \min_{\E\rightarrow \bar{\E}} I(A:B | \bar{\E}),\label{eq:intinf}\eea over all Eve's stochastic mappings
$\E\rightarrow \bar{\E}$.

In the following subsections, we show that measurements on the Smolin states in the computational basis in fact give bound information. We then show that bound information can be superactivated in a finite-copy scenario, analogously to the
quantum case.

\subsection{Bound information and the unlockability}
\label{subsec:unl}

The entanglement properties can be related to cryptographic properties of the probability distributions that are obtained by measuring  given quantum states. Without loss of generality, one can assume that Eve has access  to the rest of legitimate parties, and this is expressed by the fact that Eve holds the purification. For instance, when the Smolin state $\rho_{ABCD}$ is shared, one can find a state $|\psi\rangle_{ABCD\E}$ such that $\rho_{ABCD} = \tr_{\E}|\psi\rangle\langle\psi|_{ABCD\E}$. In this way, Eve is naturally included and her correlations with the legitimate parties are readily shown. Denoted by positive operator $M_{\alpha}$ of party $\alpha$, the probability distribution $P_{ABCD\E}$ of  the five parties reads, \bea \tr[M_{A}\otimes M_{B}\otimes M_{C}\otimes M_{D}\otimes M_{\E} |\psi\rangle\langle\psi|_{ABCD\E}].\nonumber \eea Suppose that measurements applied by the parties are in the computational basis. The probability distribution is explicitly given by,
\begin{equation}\label{smolin}
  \begin{array}{cccccc}
    \hline \hline
    A & C & B & D & \E & P_{ACBD\E} \\
    \hline
    0 & 0 & 0 & 0 & \epsilon_{1} & 1/8\\
    0 & 0 & 1 & 1 & \epsilon_{2} & 1/8\\
    1 & 1 & 0 & 0 & \epsilon_{2} & 1/8\\
    1 & 1 & 1 & 1 & \epsilon_{1} & 1/8\\
    0 & 1 & 0 & 1 & \epsilon_{3} & 1/8\\
    0 & 1 & 1 & 0 & \epsilon_{4} & 1/8\\
    1 & 0 & 0 & 1 & \epsilon_{4} & 1/8\\
    1 & 0 & 1 & 0 & \epsilon_{3} & 1/8\\
    \hline \hline
  \end{array}.
\end{equation}

In what follows, we show that, analogously to the quantum case, the distribution (\ref{smolin}) in fact contains bound information which is also unlockable.
All these are obtained as classical analogues of the properties shown in the Smolin state in Eq. (\ref{eq:smolin}).
\begin{itemize}
    \item (i') \emph{Invariance under permutations.} The distribution (\ref{smolin}) is invariant under permutations of parties, i.e. $P_{ACBD\E} = P_{ABCD\E} = P_{ADBC\E}$.
    \item (ii') \emph{Undistillability.} The distribution $P(A,C,B,D,\E)$ is undistillable in every bipartition across two parties versus the others. That is, for instance
in the bipartition between $AC$ and $BD$, it holds that \bea I (AC
: BD ~{\downarrow} ~\E ) = 0, \eea where Eve's local mapping is
given by, $\epsilon_{2}\rightarrow \epsilon_{1} $ and
$\epsilon_{3} \rightarrow \epsilon_{4}$. From the relation in
\eqref{eq:intinq}, it follows that $S(AC:BD \| \E) =0 $. Then, the
permutational invariance in (i') implies $S(AB : CD \| \E) = S(AD
: BC \| \E) = 0$, and therefore none of two parties can distill an
sbit.
    \item (iii') \emph{Unlockability.} The secret correlations existing in (\ref{smolin})
are unlockable. Suppose two parties, for instance $B$ and $D$,
join together and post-select either case that they are the same
or different. Let us now restrict to the case that $B$ and $D$
accept when they share the same bit values. Then, the distribution
is given by
\begin{equation}\label{unl}
  \begin{array}{cccccc}
    \hline \hline
    A & C & B & D & \E & P_{ACBD\E} \\
    \hline
    0 & 0 & 0 & 0 & \epsilon_{1} & 1/4\\
    0 & 0 & 1 & 1 & \epsilon_{2} & 1/4\\
    1 & 1 & 0 & 0 & \epsilon_{2} & 1/4\\
    1 & 1 & 1 & 1 & \epsilon_{1} & 1/4\\
    \hline \hline
  \end{array}.
\end{equation}
This means that an sbit is distilled between $A$ and $C$, since it
is clear in the distribution (\ref{unl}) that i) $P_{AC}(0,0) =
P_{AC}(1,1) = 1/2$ and ii) $P_{AC\E}(a,c,e) =
P_{AC}(a,c)P_{\E}(e)$. For the other case that $B$ and $D$ accept
whenever they share different bit values, applying the bit-flip
operation either $A$ and $C$, Alice and Clare can distill an sbit.
From the symmetry property in (i'), it immediately follows that
any two parties who join and collaborate to identify the shared
state can allow the other two parties to distill an sbit. As an
sbit is distilled, this also means that the probability
distribution in \eqref{smolin} consists of secret correlations.
Together with the undistillability in (ii'), it is shown that the
distribution in \eqref{smolin} indeed contains bound information.

\end{itemize}
\subsection{Superactivation}
\label{subsec:csuperact}

In this subsection, we show that bound information can be superactivated in a finite-copy scenario. We first show that an sbit can be distilled by two parties when two copies of the bound information in Eq. (\ref{smolin}) are shared by five parties. Then, it follows with more copies that multipartite sbits are distilled by the five parties.

Let us begin with the the following probability distribution shared by the five parties, \bea P_{ABCDE} = P_{A_1 C_1 B_1 D} P_{A_2 B_2 C_2 E}, \label{eq:cde}\eea where each four-partite distribution is shown in Eq. (\ref{smolin}) and the first and the second copies are labeled. Note that the distribution in Eq. (\ref{eq:cde}) can also be obtained by directly measuring the tensored state in \eqref{eq:DEactivation} in the computational basis. To be explicit, the distribution in Eq. (\ref{eq:cde}) shows, for $i,j=0,1$,
\begin{equation}\label{prob1}
  \begin{array}{cccccccccccccc}
    \hline \hline
    A_{1} & A_{2} & B_{1} & B_{2} & C_{1} & C_{2} & D & E & \E_{1} & \E_{2} \\
    \hline
    i  & j  & i  & j  & i  & j   & i  & j   & \epsilon_{1} & f_{1} \\
    i  & j  & i  & j  & i  & j+1 & i  & j+1 & \epsilon_{1} & f_{2} \\
    i  & j  & i  & j+1& i  & j   & i  & j+1 & \epsilon_{1} & f_{3} \\
    i  & j  & i  & j+1& i  & j+1 & i  & j   & \epsilon_{1} & f_{4} \\
    i  & j  & i+1& j  & i  & j   & i+1& j   & \epsilon_{2} & f_{1} \\
    i  & j  & i+1& j  & i  & j+1 & i+1& j+1 & \epsilon_{2} & f_{2} \\
    i  & j  & i+1& j+1& i  & j   & i+1& j+1 & \epsilon_{2} & f_{3} \\
    i  & j  & i+1& j+1& i  & j+1 & i+1& j   & \epsilon_{2} & f_{4} \\
    i  & j  & i  & j  & i+1& j   & i+1& j   & \epsilon_{3} & f_{1} \\
    i  & j  & i  & j  & i+1& j+1 & i+1& j+1 & \epsilon_{3} & f_{2} \\
    i  & j  & i  & j+1& i+1& j   & i+1& j+1 & \epsilon_{3} & f_{3} \\
    i  & j  & i  & j+1& i+1& j+1 & i+1& j   & \epsilon_{3} & f_{4} \\
    i  & j  & i+1& j  & i+1& j   & i  & j   & \epsilon_{4} & f_{1} \\
    i  & j  & i+1& j  & i+1& j+1 & i  & j+1 & \epsilon_{4} & f_{2} \\
    i  & j  & i+1& j+1& i+1& j   & i  & j+1 & \epsilon_{4} & f_{3} \\
    i  & j  & i+1& j+1& i+1& j+1 & i  & j   & \epsilon_{4} & f_{4} \\
    \hline \hline
  \end{array}.
\end{equation} This expression can be obtained using a simpler form of Eq. (\ref{smolin}) that is shown in the appendix A.

The classical analogue of the quantum teleportation, which is to be used in the activation protocol, is in fact the one-time pad that securely sends a classical bit by sacrificing an sbit. For convenience, we also call this "teleporting" classical bits, which works as follows. Assume that an sbit $s$, is shared by two honest parties. The sender encodes a message $x$ and publicly announces the addition $(x+s)$, so that the receiver can decode the message by adding the shared sbit, $(x+s)+s$. Since the value of the sbit $s$ is not known to anyone else, one can only guess a random bit from the public communication.

The activation protocol is obtained by translating the quantum one, and works as follows, see also Fig. \ref{fig1}. First, Alice teleports her bit in $A_{2}$ to Clare, sacrificing an sbit in $A_{1}C_{1}$. Clare then has a new value in the register, $C_{1}^{'} = C_{1} + A_{1} + A_{2}$, and the probability distribution becomes as follows, \begin{equation}\label{prob2}
  \begin{array}{cccccccccccccc}
    \hline \hline
    B_{1} & B_{2} &  C_{1}^{'} & C_{2} & D & E & \E_{1} & \E_{2} \\
    \hline
    i  & j  & j  & j   & i  & j   & \epsilon_{1} & f_{1} \\
    i  & j  & j  & j+1 & i  & j+1 & \epsilon_{1} & f_{2} \\
    i  & j+1& j  & j   & i  & j+1 & \epsilon_{1} & f_{3} \\
    i  & j+1& j  & j+1 & i  & j   & \epsilon_{1} & f_{4} \\
    i+1& j  & j  & j   & i+1& j   & \epsilon_{2} & f_{1} \\
    i+1& j  & j  & j+1 & i+1& j+1 & \epsilon_{2} & f_{2} \\
    i+1& j+1& j  & j   & i+1& j+1 & \epsilon_{2} & f_{3} \\
    i+1& j+1& j  & j+1 & i+1& j   & \epsilon_{2} & f_{4} \\
    i  & j  & j+1& j   & i+1& j   & \epsilon_{3} & f_{1} \\
    i  & j  & j+1& j+1 & i+1& j+1 & \epsilon_{3} & f_{2} \\
    i  & j+1& j+1& j   & i+1& j+1 & \epsilon_{3} & f_{3} \\
    i  & j+1& j+1& j+1 & i+1& j   & \epsilon_{3} & f_{4} \\
    i+1& j  & j+1& j   & i  & j   & \epsilon_{4} & f_{1} \\
    i+1& j  & j+1& j+1 & i  & j+1 & \epsilon_{4} & f_{2} \\
    i+1& j+1& j+1& j   & i  & j+1 & \epsilon_{4} & f_{3} \\
    i+1& j+1& j+1& j+1 & i  & j   & \epsilon_{4} & f_{4} \\
    \hline \hline
  \end{array}.
\end{equation}
Next, Bob teleports his value in $B_{2}$ to David sacrificing an sbit in $B_{2}D$. Then, David holds a new value $D' = D+B_{1}+B_{2}$, with which the probability distribution of the four parties is given by
\begin{equation}\label{sprob3}
  \begin{array}{cccccccccccccc}
    \hline \hline
    C_{1}^{'} & C_{2} & D^{'} & E & \E_{1} & \E_{2} \\
    \hline
    j  & j   & j  & j   & \epsilon_{m} & f_{1} \\
    j  & j+1 & j  & j+1 & \epsilon_{m} & f_{2} \\
    j  & j   & j+1& j+1 & \epsilon_{m} & f_{3} \\
    j  & j+1 & j+1& j   & \epsilon_{m} & f_{4} \\
    \hline \hline
    \end{array}
\end{equation} where $m=1,2,3,4$. The explicit form of the distribution of Eq. (\ref{sprob3}) is shown in the appendix B. Now, the distribution in Eq. \eqref{sprob3} is identical to the bound information in Eq. \eqref{smolin}. Remind that the secret correlations in \eqref{smolin} is unlockable, as it was shown in Sec.\ref{subsec:unl}. Therefore, Clare, who is with two bits $C_{1}^{'}$ and $C_{2}$, announces if her two values are the same or not, depending on which, by applying local operations David and Elena can share an sbit: if it is announced that $C_{1}^{'}$ and $C_{2}$ are unequal, either David or Elena applies the bit-flip operation. It is therefore shown that an sbit can be distilled between $D$ and $E$.

Moreover, symmetrizing the distribution in Eq. (\ref{eq:cde}), i.e., \bea P_{ABCD\E_{1}} P_{ABCE\E_{2}} P_{ABDE \E_{3}} P_{ACDE \E_{4}} P_{BCDE \E_{5}}, \label{dprob}\eea any two parties among the five can distill sbits against an eavesdropper who holds the five random variables $\E_{1}\E_{2}\E_{3}\E_{4}\E_{5}$. Therefore, it is straightforward that, with more copies, the five parties can share secrecy.

\section{Distribution of entanglement and secrecy}
\label{sec:dis}

In this section, we show a usefulness of undistillable correlations in quantum and classical scenarios, respectively, namely that they can be used to distribute multipartite distillable correlations. In the quantum scenario, we consider distribution of multi-partite GHZ state, \bea |\phi_{N}\rangle = (|0\rangle^{\otimes N} + |1\rangle^{\otimes N})/\sqrt{2}. \nonumber \eea We show that tripartite GHZ state can be deterministically extended into four parties using LOCC when the Smolin state is shared by the four parties.

We also derive a classical analogue of the quantum state distribution. Multipartite sbits of $N$ parties, say $A_{1},\cdots,A_{N}$, is a classical analogue of the $N$-partite GHZ state, being defined as the following probability distribution \bea P_{A_{1},\cdots,A_{N}}(a_{1},\cdots,a_{N}) &=& \delta_{a_{1},a_{2}}\delta_{a_{2},a_{3}}\cdots \delta_{a_{N-1},a_{N}}/2, \nonumber\\
P_{A_{1},\cdots,A_{N},\E}(a_{1},\cdots,e) &=& P_{A_{1},\cdots,A_{N}}(a_{1},\cdots,a_{N})P_{\E}(e).\nonumber\eea  We then show that the tripartite sbit  can be extended into four parties using LOPC when the bound information in Eq. (\ref{smolin}) is shard by the four parties. Note that in both quantum and classical scenarios the distribution scheme works deterministically.

\subsection{Quantum scenario}
\label{subsec:qdis}

Suppose that Alice, Bob, Clare, and David share the Smolin state, and that only three of them, say Alice, Bob, and Clare, additionally share a tripartite GHZ state as follows \bea  \mu_{ABCD} = |\eta\rangle\langle\eta|_{ABCD} \otimes \rho_{ABCD} \label{eq:twostate}\eea where  $|\eta\rangle_{ABCD} = |\phi_{3}\rangle_{ABC} \otimes|+\rangle_{D}$ and $ |+\rangle_{D} = (|0\rangle + |1\rangle)/\sqrt{2}$. Let $\Lambda_{\alpha}$ for $\alpha=A,B,C,D$ denote the local operation performed by the party $\alpha$. The goal is now to show that the state $\mu_{ABCD}$ can be transformed to $|\phi_{4}\rangle$ using some local operations $\Lambda_{\alpha}$. To this end, the local operation, $\Lambda_{\alpha}:\C^{2}\otimes \C^{2} \rightarrow \C^{2}$, mapping from two-qubit to a single qubit states, can be explicitly constructed in terms of the Kraus operators, $K_{0}^{\alpha} = |0\rangle\langle 00| + |1\rangle\langle 11|$ and $K_{1}^{\alpha} = |0\rangle\langle 01| + |1\rangle\langle 10|$ as follows \bea \Lambda_{\alpha}(\cdot) = \sum_{i=0,1} K_{i}^{\alpha}(\cdot)K_{i}^{\alpha\dagger}. \label{eq:map}\eea The above can be in fact rephrased as collective measurement by which it can only be known if two qubit systems are in the same state or not, leaving a single qubit system.

Now, the four parties apply the local operation (\ref{eq:map}) to the state in (\ref{eq:twostate}). Suppose that four parties get measurement outcomes $(i_{A},j_{B},k_{B},l_{D})$. This happens with probability \bea \tr[\mu_{ABCD} K_{i_{A}}^{A\dagger}K_{i_{A}}^{A}\otimes K_{j_{B}}^{B\dagger}K_{j_{B}}^{B } \otimes K_{k_{C}}^{C\dagger} K_{k_{C}}^{C} \otimes K_{l_{D}}^{D\dagger} K_{l_{D}}^{D}],\nonumber\eea and the state resulted in the four parties is, \bea |\phi^{v}\rangle = \Id_{A} \otimes \Id_{B} \otimes\Id_{C} \otimes (\sigma_{D}^{x})^{v}|\phi_{4}\rangle, \eea where $v=i_{A}+j_{B}+k_{B}+l_{D}$ and $\sigma_{D}^{x}$ denotes the Pauli matrix $\sigma_x$ in the David's side. Therefore, using classical communication to discuss the measurement outcomes, the four parties can compute, $v = i_{A} + j_{B}  + k_{C} + l_{D}$. If $v$ is an even number, this means that the four-partite GHZ state is already shared. Otherwise, David applies the $\sigma_{x}$ operation to his qubit, and the four-partite GHZ state can be shared.

\begin{figure}
\begin{center}  
  \includegraphics[width= 7 cm]{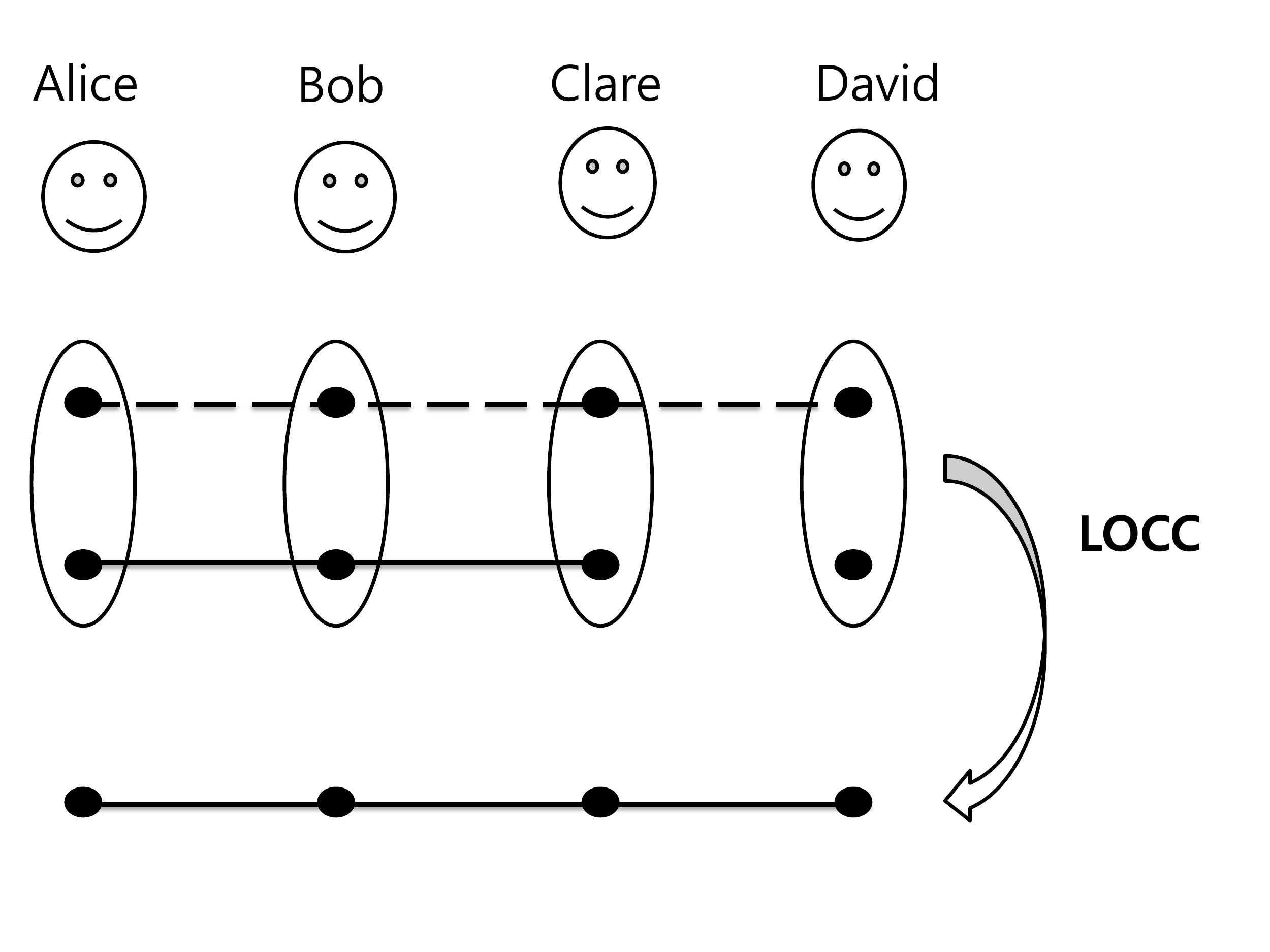}
  \caption{Distribution of multipartite pure entanglement and secret key: Four parties share the Smolin states (dashed line) and are allowed to apply LOCC. Then, the tripartite GHZ state (solid line) can be distributed over the four parties using LOCC. The classical analogue also follows: the tripartite sbit can be distributed over the four using LOPC when the four-partite bound information in Eq. (\ref{smolin}) is shared.}\label{fig2}
\end{center}
\end{figure}

\subsection{Classical scenario}
\label{subsec:cdis}

Suppose that the four parties share the bound information in (\ref{smolin}) and that only three of them, say Alice, Bob, and Clare, share an sbit in unknown value $s$ for $s\in\{0,1\}$. Let $i_{k}$ for $k=A,B,C,D$ denote the value of the party $k$. The goal is then to distribute the tripartite sbit into the four parties using LOPC, such that David also securely share the sbit with the rest.

The distribution protocol works as follows, see also Fig. \ref{fig2}. Each of the three parties sharing the sbit, individually and locally copy the sbit and then compute the parity of the two bits, one from the sbit $s$ and the other from the bound information $i_{k}$ for $k=A,B,C$. Then, using the distribution in (\ref{smolin1}) which is a simpler expression of (\ref{smolin}), the distribution is transformed as follows,
\begin{equation}\label{eq:csec}
  \begin{array}{cccccc}
    \hline \hline
    A & C & B & D & \E & P_{ABCD\E} \\
    \hline
    s+i & s+i & s+i & i & \epsilon_{1} & 1/8 \\
    s+i & s+i & s+i+1 & i+1 & \epsilon_{2} & 1/8\\
    s+i & s+i+1 & s+i & i+1 & \epsilon_{3} & 1/8\\
    s+i & s+i+1 & s+i+1 & i & \epsilon_{4} & 1/8\\
    \hline \hline
  \end{array}.
\end{equation}
Afterwards, each party publicly announces the parity bit $s+i_{k}$, so that David knows them and computes the sum of the announced bit values, denoted by $v_{D} = \sum_{k=A,B,C} (s+i_{k})$. David then adds $v_{D}$ to his bit $i_{D}$ as follows,
\begin{equation}\label{eq:csec2}
  \begin{array}{cccccc}
    \hline \hline
    A & C & B & D & \E & P_{ABCD\E} \\
    \hline
    s+i & s+i & s+i & i + v_{D} & \epsilon_{1} & 1/8 \\
    s+i & s+i & s+i+1 & i+1 + v_{D} & \epsilon_{2} & 1/8\\
    s+i & s+i+1 & s+i & i+1 + v_{D} & \epsilon_{3} & 1/8\\
    s+i & s+i+1 & s+i+1 & i + v_{D} & \epsilon_{4} & 1/8\\
    \hline \hline.
  \end{array}
\end{equation}

Eve has classical bits that are correlated with the four parties as it is shown in Eq. (\ref{smolin}), and also listens to the announced bit values $s+i_{k}$ from the public communication. Then, as it is shown in Eq. (\ref{eq:csec}), Eve can only discriminates among the four possibilities of $\epsilon_i$ for $i=1,2,3,4$. However, the sbit $s$ shard by the three parties has not been known to Eve, who can make a random guess.

As it is shown in (\ref{eq:csec2}), the bit value of David results in, and is explicitly computed as \bea i_{D} + \sum_{k=A,B,C} s+i_{k} =s. \label{sum} \eea This is because, from the distribution of the bound information (\ref{smolin}), it holds that $\sum_{k=A,B,C,D} i_{k} = 0$. Hence, it is shown that a multipartite sbit can be distributed securely via bound information together with LOPC.

\section{Conclusion}
\label{conlusion}

We have shown a case of four-partite bound information and its properties, unlockability and superactivation. All these are obtained by deriving classical analogues of the Smolin state and its quantum effects, super-activation and unlockability in bound entangled states. It would be interesting to investigate which properties of quantum correlations can or cannot have their classical counterparts. For instance, existence of bipartite bound information remains open and is an challenging issue. Finally, we have shown a usefulness of undistillable correlations: bound entanglement and bound information can be used to distribute a multipartite GHZ state and multipartite sbits in quantum and classical scenarios, respectively.

\section*{Acknowledgement}
We are grateful to A. Ac\'in for helpful discussions and comments.
This work is supported by Consolider-Ingenio QOIT projects and the Korea Research Foundation Grant, KRF-2008-313-C00185. J.B. also thanks the Institut Mittag-Leffler
(Djursholm, Sweden) for the support during his visit.

\section*{Appendix A: Derivation of \eqref{prob1} }
\label{appendixa}

By individual measurement to each copy of two Smolin states in \eqref{eq:DEactivation}, the five parties share measurement data such that Alice, Bob, and Clare possess two values labeled $1$ and $2$ and David and Elena keep single values. Both the first and the second distributions in the form in (\ref{smolin}) can be written in a simpler  form as follows. For the first copy, \begin{equation}\label{smolin1}
  \begin{array}{cccccc}
    \hline \hline
    A_{1} & C_{1} & B_{1} & D_{1} & \E_{1} & P_{A_{1}B_{1}C_{1}D\E} \\
    \hline
    i & i & i & i & \epsilon_{1} & 1/8 \\
    i & i & i+1 & i+1 & \epsilon_{2} & 1/8\\
    i & i+1 & i & i+1 & \epsilon_{3} & 1/8\\
    i & i+1 & i+1 & i & \epsilon_{4} & 1/8\\
    \hline \hline
  \end{array}
\end{equation} where $i=0,1$, and for the second copy of $A_{2}$, $B_{2}$, $C_{2}$ and $E$, assuming Eve holding the second parameter $f_{k}$, $k=1,2,3,4$,
\begin{equation}\label{smolin2}
  \begin{array}{cccccc}
    \hline \hline
    A_{2} & B_{2} & C_{2} & E & \E_{2} & P_{A_{2}B_{2}C_{2}D\E} \\
    \hline
    j & j & j & j & f_{1} & 1/8\\
    j & j & j+1 & j+1 & f_{2} & 1/8\\
    j & j+1 & j & j+1 & f_{3} & 1/8\\
    j & j+1 & j+1 & j & f_{4} & 1/8\\
    \hline \hline
  \end{array}
\end{equation} for $j=1,2$. The full probability obtained by measuring the state in (\ref{eq:DEactivation}) is then shown in (\ref{prob1}).

\section*{Appendix B: The full distribution of \eqref{sprob3}}
\label{appendixb}

The full distribution of (\ref{sprob3}) is explicitly shown as follows, for different values of $\E_{1}$,

\begin{equation}\label{prob3}
  \begin{array}{cccccccccccccc}
    \hline \hline
    C_{1}^{'} & C_{2} & D^{'} & E & \E_{1} & \E_{2} \\
    \hline
j  & j   & j  & j   & \epsilon_{1} & f_{1} \\
j  & j+1 & j  & j+1 & \epsilon_{1} & f_{2} \\
j  & j   & j+1& j+1 & \epsilon_{1} & f_{3} \\
j  & j+1 & j+1& j   & \epsilon_{1} & f_{4} \\
j  & j   & j  & j   & \epsilon_{2} & f_{1} \\
j  & j+1 & j  & j+1 & \epsilon_{2} & f_{2} \\
j  & j   & j+1& j+1 & \epsilon_{2} & f_{3} \\
j  & j+1 & j+1& j   & \epsilon_{2} & f_{4} \\
j+1& j   & j+1& j   & \epsilon_{3} & f_{1} \\
j+1& j+1 & j+1& j+1 & \epsilon_{3} & f_{2} \\
j+1& j   & j  & j+1 & \epsilon_{3} & f_{3} \\
j+1& j+1 & j  & j   & \epsilon_{3} & f_{4} \\
j+1& j   & j+1& j   & \epsilon_{4} & f_{1} \\
j+1& j+1 & j+1& j+1 & \epsilon_{4} & f_{2} \\
j+1& j   & j  & j+1 & \epsilon_{4} & f_{3} \\
j+1& j+1 & j  & j   & \epsilon_{4} & f_{4} \\
    \hline \hline
  \end{array}.
\end{equation}

For cases when Eve is with $\epsilon_{3}$ or $\epsilon_{4}$, the distribution in (\ref{sprob3}) can be obtained by replacing $j$ with $j+1$ in (\ref{prob3}).

\end{document}